\documentclass[%
 aps,
 prl,
 longbibliography,
 sd,%
 amsmath,amssymb,
 reprint,%
]{revtex4-1}

\usepackage[dvipdfmx]{graphicx}
\usepackage{dcolumn}
\usepackage{bm}

\usepackage{setspace,color,url}
\usepackage{comment}
\usepackage{here}
\usepackage{subfigure}
\usepackage{balance}
\usepackage{color}
\usepackage{ulem}
\newcommand{\rev}[1]{{\color{black}#1}}

\begin{document}

\preprint{Submitted to Physical Review **}

\title{Electro-osmotic diode based on colloidal nano-valves between double membranes}
%
\author{Shihori Koyama}
\affiliation{Toyota Central R\&D Labs., Inc., Bunkyo-ku, Tokyo 112-0004, Japan}
\author{Daisuke Inoue}
\affiliation{Toyota Central R\&D Labs., Inc., Bunkyo-ku, Tokyo 112-0004, Japan}
\author{Akihisa Okada}
\affiliation{Toyota Central R\&D Labs., Inc., Bunkyo-ku, Tokyo 112-0004, Japan}
\author{Hiroaki Yoshida}
\email{h-yoshida@mosk.tytlabs.co.jp}
\affiliation{Toyota Central R\&D Labs., Inc., Bunkyo-ku, Tokyo 112-0004, Japan}
\date{\today}
%
\begin{abstract}
The rectification of electro-osmotic flows is important in micro/nano fluidics applications such as micro-pumps and energy conversion devices.
Here, we propose a simple electro-osmotic diode in which colloidal particles are contained between two parallel membranes with different pore densities.
While the flow in the forward direction just pushes the colloidal particles 
toward the high-pore-density membrane, the backward flow is blocked by the particles near the low-pore-density membrane, which clog the pores.
Nonequilibrium molecular dynamics simulations show a
strong nonlinear dependence on the electric field
for both the electric current and electro-osmotic flow,
indicating diode characteristics.
A mathematical model to reproduce the electro-osmotic diode behavior is constructed, introducing an effective pore diameter as a model for pores clogged by the colloidal particles. Good agreement is obtained between the proposed model with estimated parameter values and the results of direct molecular dynamics simulations. 
The proposed electro-osmotic diode has potential application in downsized microfluidic pumps, e.g., the pump induced under AC electric fields.
\end{abstract}
\maketitle
%
%
%
%
\section{\label{sec_intro}Introduction}

Micro- and nano-scale transport systems have attracted interest for application in fields ranging from bio- and medical-technologies to new energy conversion and storage~\cite{DYS+2004,LPG+2005,BPF+2006,DM2006,VBS+2006,FOL+2010,SPB+2013,KKH+2017,UKK+2020}. Various devices exploiting the phenomena particular to the small scale flows have been proposed~\cite{LS2004,SSA2004,SQ2005}.
In typical micro- and nano-fluidic devices, an external field, such as an electrical or chemical potential field, applied to an electric double layer formed in the vicinity of a solid-liquid interface, induces electrokinetic phenomena, which are typically accompanied by interfacial liquid flow. The high surface area to volume ratio of micro- and nano-scale systems enhances these phenomena, which have thus been studied for application as the driving force in small-scale transport systems~\cite{SHR2008,BC2010B}.

A representative electrokinetic phenmenon is the electro-osmotic flow, which is induced by an external electrical field~\cite{PHS1974,STZ+2004,LD2006,KWL+2007,NCH+2015}.
Since the flow is directly induced by the electric field, i.e., there are no moving parts, it is recognized as a key technique for downsizing of pumps.
Applications of porous materials, e.g. a glass porous material, as a source of many pores were extensively studied as a promising setup for downsized pumps~\cite{YS2003,YHZ+2003,VXM+2007}.
The use of small size fabrication techniques such as microlithography and chemical etching techniques on a substrate was alternatively studied~\cite{UTL+2006,HC2007}.
Recently, all-plastic nano fabrication of electro-osmotic flow membrane was  reported~\cite{BR2017}. Along with these experiments, theoretical analyses for predicting the flow rate~\cite{WLC2008,MSG2014,SMG2014,MHG2017} and molecular dynamics (MD) simulations~\cite{QA2004,CNW+2008,RP2013,YMK+2014} have also been conducted for nano sized systems.

One of the challenges for the practical application of electro-osmotic flow is
that the long-term application of a DC voltage as an external field electrolyzes water. To prevent the electrolysis of solvent water, many attempts have been made by using, e.g., traveling-wave potentials~\cite{RMG+2005}, 
three-dimensional stepped electrode arrays~\cite{HBT2010}, ratcheted electrodes~\cite{SS2019}, and an AC voltage whose period is shorter than a characteristic time scale of electrolysis~\cite{WRM2016,LWH+2018}.
In these methods, rectifying the forward and backward electro-osmotic flows is commonly a key ingredient, to induce a net one-way flow.

The rectification and control of flows have been extensively studied~\cite{PHS1974,DMS+1998,TMS+2006,LPA+2009,YAM2010,GMM2011,PMT+2017,MB2017,SSF+2018,NJT+2018,LAS2019,NIK+2019,HKT2020,LTM+2020} for various types of small scale systems.
Particularly, interest in rectification at the nano-scale has rapidly grown.
Examples of nano-scale rectification devices include those with membranes that have asymmetric pores, such as conical pores ~\cite{SAD+2003,SAB+2003,SKF+2005,Siwy2006,CG2010,LWH+2018,JPS+2018,BVM2020}, 
nanopipettes~\cite{UPW+2006,SB2013,DTS+2014,LWK+2015,BLY+2020}, and nanotubes~\cite{SHH+1997,SHH+2004}.
Other asymmetries have also been considered, including those caused by the asymmetric distribution of surface charges at a solid-liquid interface~\cite{DOS2005,KDC+2007,VS2007,CG2009,PGJ+2013,PZF+2015},
directed applied voltage~\cite{GFR2011,WWV+2012,WWB+2013}, and combinations of nano-porous media and ion-exchange membranes~\cite{YLZ+2017}.
In addition, the classical valve structure has been adapted for nano-scale systems~\cite{Tesar2008}. This approach is promising because a large difference between the forward and backward flows is mechanically produced. However, complex nano-fabrication methods and mechanical malfunction of sensitive structures hinder adoption.
A system with a simpler structure that spontaneously and mechanically prevents backward flow, such as that for micro-scale rectification based on active filter clogging~\cite{MLM2012,YKL+2016},
would efficiently rectify nano-scale electro-osmotic flow, expanding possible applications.

In the present study, we propose an electro-osmotic diode made of two electro-osmotic flow membranes, that quasi-mechanically suppress backward flow when an electric field is applied in the backward direction.
More specifically, the system consists of two membranes with different pore densities and colloidal particles in between (see Fig.~\ref{fig:problem}(a)).
When an electric field is applied in the forward direction, the colloidal particles approach the high-pore-density membrane, which has little impact on the generated electro-osmotic flow.
The other membrane has much fewer pores, so when an electric field is applied in the opposite direction, the colloidal particles plug the pores, blocking the backward electro-osmotic flow.
This simple diode system highlights potential application in creating a one-way flow using the electro-osmotic flow under an AC electric field as mentioned above, if the typical period is sufficiently longer than the relaxation time of the colloid motion.

We conduct MD simulations of the proposed system to demonstrate that both the electric current and electro-osmotic flow rate exhibit strong nonlinearity with respect to the electric field, to show a performance as an electro-osmotic diode.
In order to confirm the observation, next we construct a mathematical model of the electro-osmotic diode. Here we introduce a new parameter, what we call an effective pore diameter, which changes depending on the existence probability of colloidal particles near pores.
The theoretical model for electro-osmotic flow through a cylindrical pore of a finite length is extended using the effective pore diameter. 
The existence probability density functions of colloidal particles near pores are expressed in terms of the solutions of the Fokker-Planck equations for the motion of colloidal particles.
The parameters in the model equation are estimated using independent MD simulations, which are dedicated to measure the properties of colloids. The proposed model with these estimated parameters is shown to reproduce the simulation results for the entire system.

%
%
%
\section{\label{sec:problem}Setup and simulation}
\subsection{Electro-osmotic diode}
\begin{figure}[t]
 \includegraphics[width=0.95\hsize]{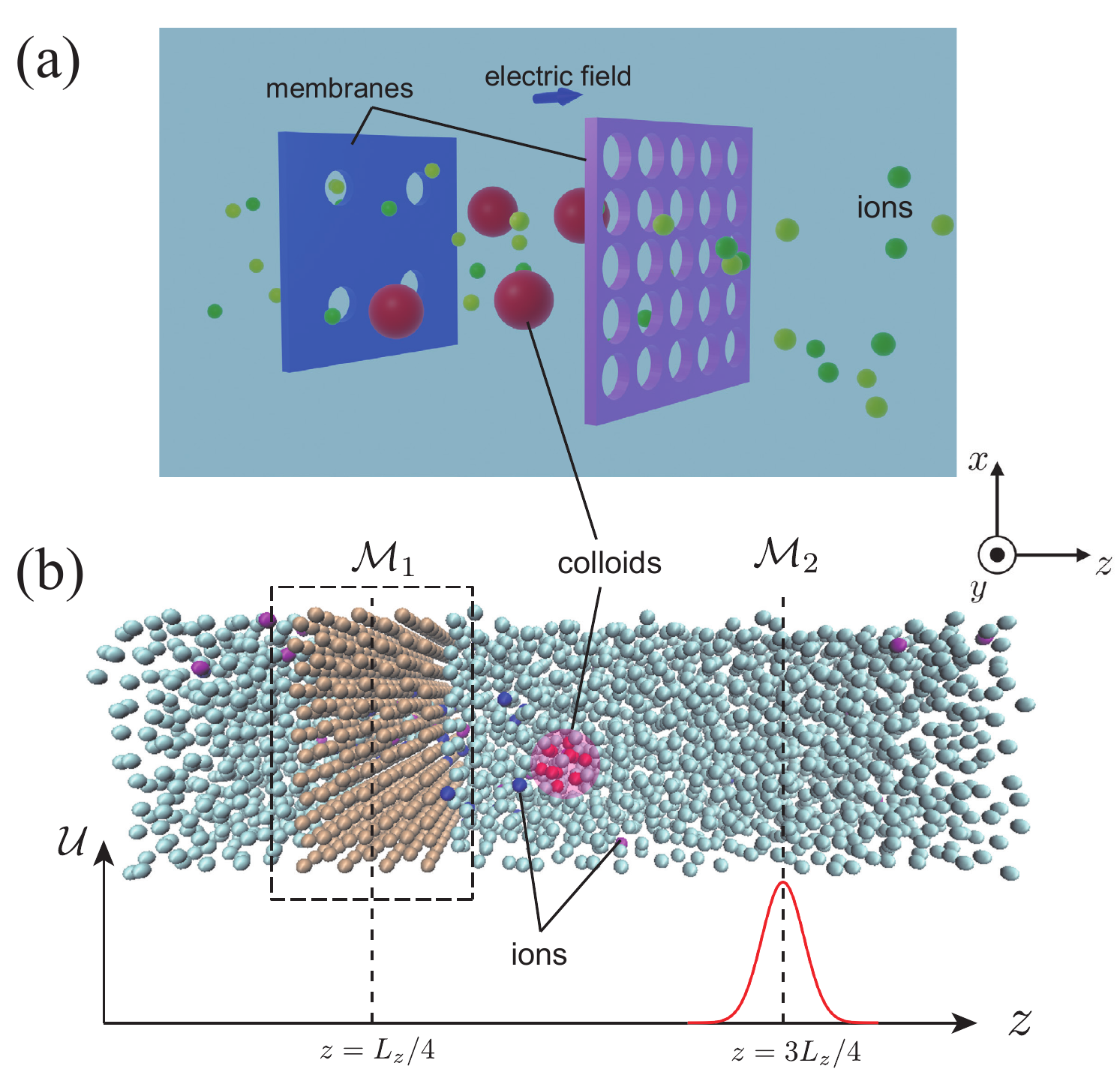}
 \caption{Electro-osmotic diode system. (a) An illustration, and (b) the model system considered in the simulations. 
 In the model system, 
 membrane $\mathcal{M}_1$ has a physical pore, and membrane $\mathcal{M}_2$ is modeled as an energy barrier that affects only the colloid particle.}
 \label{fig:problem}
\end{figure}

Let us consider a system of two membranes $\mathcal{M}_1$ and $\mathcal{M}_2$,
placed in parallel in the direction perpendicular to the $z$ axis,
and colloidal particles between the membranes. The particle size is slightly larger than the pore diameter, $D$, in the membranes. 
As illustrated schematically in Fig.~\ref{fig:problem}(a),
the number of pores for membrane $\mathcal{M}_1$ is comparable to that of the colloidal particles, whereas that for membrane $\mathcal{M}_2$ is higher.
The colloidal particles are positively charged, with a charge $q$ per particle, and the walls of the pores are negatively charged. 
If an electric field $E_z$ in the $z$ direction is applied from $\mathcal{M}_1$ to $\mathcal{M}_2$, the colloidal particles are induced to flow by electro-phoresis toward membrane $\mathcal{M}_2$. An electric field applied in the opposite direction would drive the colloidal particles toward $\mathcal{M}_1$. In the latter case, we expect that the current and electro-osmotic flow induced by $E_z$ $(<0)$ will be blocked by particles near pores.

The systems considered here for the simulations and theories are on the scale of nm $\sim$ tens of nm. Corresponding experimental setups could be constructed using materials such as polycarbonate~\cite{BR2017}, PET~\cite{WWZ+2012,WRM2016}, and carbon~\cite{MYM2001,SMN+2016} for the membranes. The track-etching technique~\cite{MJB2020,HFS2007} is able to control the size and the number of pores, as done for creating nanofilteration membranes for molecular sieving.
The electro-osmotic flow is then created using the aqueous electrolyte solution as NaCl solution.
Nano-particles such as gold nanoparticles~\cite{MCS+2004}, silica particles~\cite{VSD2004}, and hydrous zirconia particles~\cite{TSB+2001} are possible candidates for the colloidal particles.

In the present study, to focus on the interaction between the pore in membrane $\mathcal{M}_1$ and the colloidal particles, the model system shown in Fig.~\ref{fig:problem}(b) is considered
as an abstracted system for analysis, where membrane $\mathcal{M}_1$ with thickness $L$ has a single pore with a diameter $D$, and a single colloidal particle exists in the solution. Membrane $\mathcal{M}_2$ is modeled as a completely semi-permeable membrane with a virtual energy barrier that affects only colloidal particles,
i.e., the colloidal particles are completely rejected by $\mathcal{M}_2$ whereas the solvent freely passes through. Assuming periodic boundary conditions in the $x$ and $y$ directions, we investigate the behavior of the current density and electro-osmotic flow of this system using MD simulations.

%
%
\subsection{Molecular dynamics simulations}
In this study, MD simulations are performed 
with the membranes, colloidal particle, and solvent modeled using
particles. The Lennard-Jones (LJ) potential is used for inter-particle interactions.
The potential energy for the pairwise interaction is 
$\mathcal{E} = 4\epsilon ((\sigma/r_*)^{12} -  (\sigma/r_*)^{6}))$
for $r_*<r_c$, and zero otherwise.
Here, $r_*$ is the distance between the center of the interacting particles, and $\epsilon$ and $\sigma$ are parameters corresponding to the potential depth and diameter of the particles, respectively; $r_c$ is the cutoff parameter.
In the present study, common values of LJ parameters, namely $\epsilon_0$ and $\sigma_0$, are assigned to all the LJ particles, and the cutoff distance is $r_c=2.5\sigma_0$. 

Throughout the paper, the physical quantities are scaled using the following characteristic quantities: length $\sigma_0$, mass $m_0$, time $\tau_0 = \sigma_0 \sqrt{m_0/\epsilon_0}$, energy $\epsilon_0$, temperature $T_0=\epsilon_0/k_B$, charge $q_0=\sqrt{4\pi \varepsilon_0 \sigma_0 \epsilon_0}$, and electric field $E_0=\epsilon_0/\sqrt{4\pi \varepsilon_0\sigma_0^3 \epsilon_0}$, where $k_B$ is the Boltzmann constant and $\varepsilon_0$ is the permittivity of vacuum.

A snapshot of the MD simulation, illustrating the computational system, is shown in Fig.~\ref{fig:problem}(b). 
We use a seven-layer hexagonal close-packed structure stacked in the $z$ direction as membrane $\mathcal{M}_1$, the middle layer of which is placed at $z=L_z/4$, with $L_z$ being the system size in the $z$ direction. A pore at the center of the membrane is created by removing particles less than $1.2\sigma_0$ away from the center line. The uniformly distributing $12$ particles on the pore wall are monovalent anions, where the charge of each anion is $-q_0$.
Membrane $\mathcal{M}_2$, which is a semi-permeable membrane, is modeled by a virtual energy barrier that affects only the colloidal particle. 
Specifically, we assume the following Gaussian potential barrier:
\begin{align}
\mathcal{U} = U_0 \exp \left[-a_0(z-z_0)^2\right],\,\,\,z<\sigma_0,
\end{align}
where the height of barrier $U_0$ is set to $30\,\epsilon_0$, which is sufficiently large to prevent the colloidal particle from passing through $\mathcal{M}_1$.
The center of the barrier is at $z_0=3L_z/4$ and the parameter $a_0$, which determines the potential width, is $10\sigma_0^{-2}$.

The colloidal particle is composed of $12$ cations located at the vertices of an icosahedron with a diagonal length of $2\,\sigma_0$.
The net charge of the colloidal particle is $q$, i.e., the charge of each ion is $q/12$.
The electrolyte solution consists of $2410$ particles, including $20$ monovalent anions, and $32-q$ monovalent cations,
where the number of cations is determined such that the total charge in the system (including the surface charge on the pore wall) is zero. 

We use the open-source package LAMMPS~\cite{lammps} for the MD simulations. The velocity Verlet method is employed for the time integration of the Newton equation for each particle.
To deal with the long-range Coulomb interaction, we employ the particle-particle-particle-mesh (PPPM) method~\cite{HE1988}. 
A periodic boundary condition is applied in all directions and the NVT ensemble is used to maintain the temperature of the system at $T_0$. The time step is $dt=0.002$\,$\tau_0$.
The typical size of the system is $L_x=L_y=10$\,$\sigma_0$ in the $x$ and $y$ directions and $L_z$ is $L_z\sim 40$\,$\sigma_0$.
The precise value of $L_z$ is determined such that the pressure in the $z$ direction is $1.0 \pm 0.1$\,$\epsilon_0/\sigma_0^3$. 

\begin{figure}[t]
 \includegraphics[width=1\hsize]{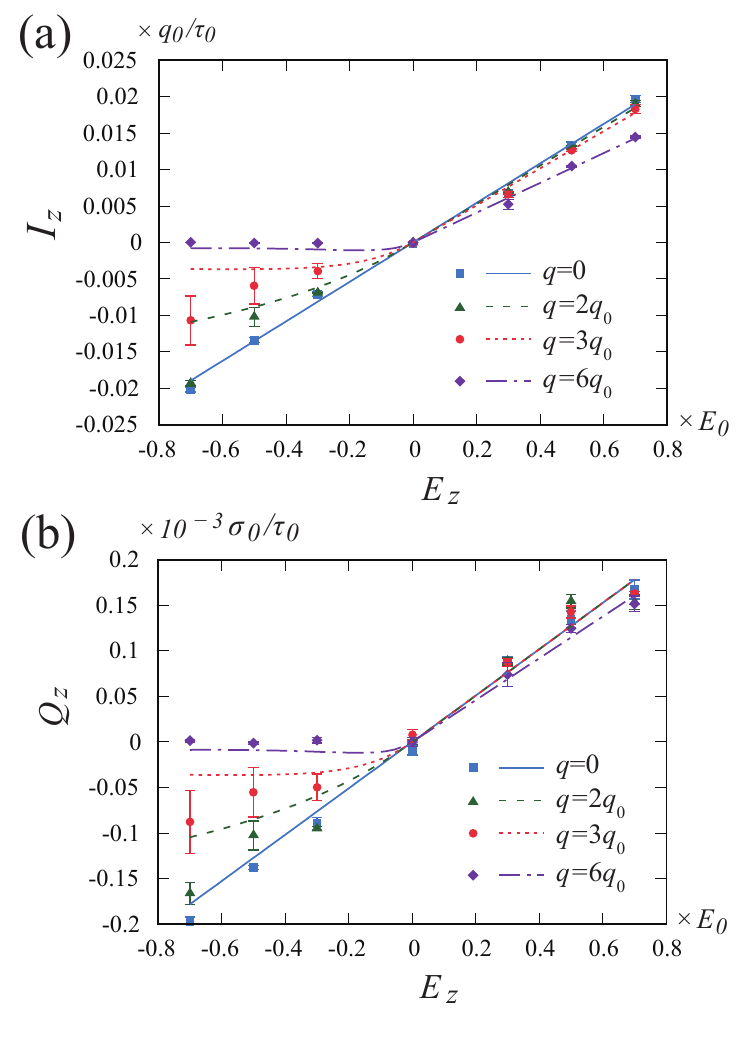}
 \caption{
 Nonlinear response of electro-osmotic diode obtained using MD simulations.
 (a) Current $I_z$ and (b) electro-osmotic flow rate $Q_z$ as functions of electric field $E_z$ for various values of colloidal particle charge $q$. 
 The lines are the corresponding results obtained from the model described in Secs.~\ref{sec:model} and \ref{sec:discussion}.
 The error bars indicate the standard deviation for the data obtained for different initial configurations.
 }
 \label{fig:result}
\end{figure}

The simulation results for various values of colloidal particle charge $q$ are shown in Fig.~\ref{fig:result}. The current $I_z$ and the electro-osmotic flow rate $Q_z$ are plotted as functions of the electrical field $E_z$ in the range $-0.7E_0\le E_z\le 0.7E_0$.
To obtain these results, the system is first equilibrated with no electric field for $10^5$ time steps before the production run, with the pressure kept at $\epsilon_0/\sigma_0^3$ and the temperature kept at $T_0$. Then the electric field is turned on and the production run is carried out for more than $10^7$ steps. The current  and flow rate at a time instance are calculated 
as $I_z=(S/V)\sum_\mathrm{ions}q_iv_{zi}$ and $Q_z=(1/N)\sum_\mathrm{solution}v_{zi}$,
where $q_i$ and $v_{zi}$ are the charge and velocity in the $z$ direction of a particle, $S$ and $V$ are respectively the area in the $x$-$y$ plane and the volume 
of the measured region, and $N$ is the number of the measured particles.
These instant values are averaged over the last $8\times 10^6$ time steps of the production run.
We perform three production runs for each case using different initial configurations.
Each point in Fig.~\ref{fig:result} is the mean of three values and the lines are the results of the model (see Sec.~\ref{sec:model} for details). 

\begin{figure}[tb]
 \centering
 \includegraphics[width=0.95\hsize]{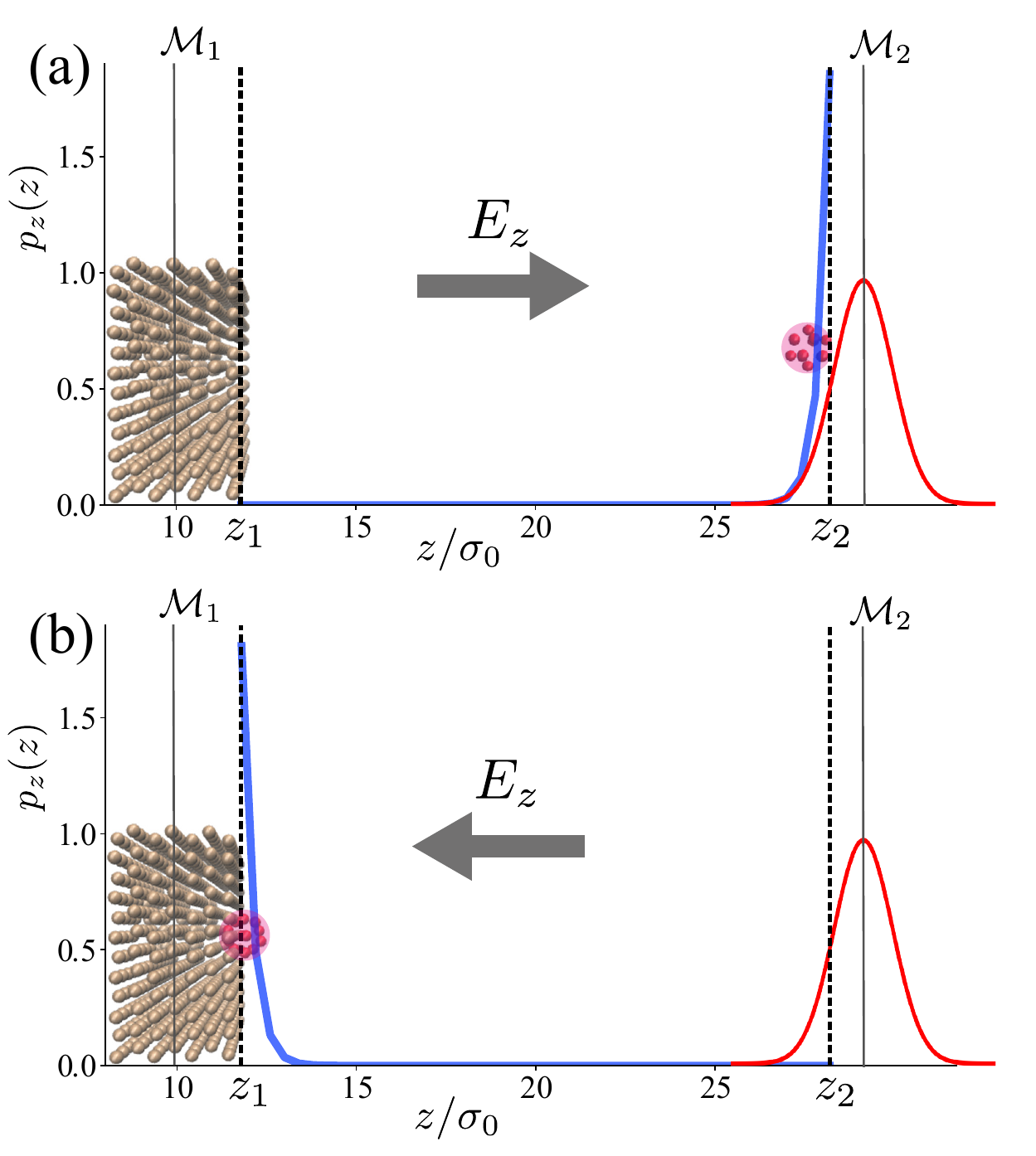}
 \caption{
 Behavior of colloidal particle under electric field applied in (a)forward ($E_z=0.7E_0$) and (b)backward ($E_z=-0.7E_0$) directions. The horizontal axis represents the position along the $z$-axis and the vertical axis is the existence probability of the colloidal particle in the $z$ direction $p_z(z)$ (blue solid line; see Sec.~\ref{sec:discussion} for details).
 }
 \label{fig:md_simulation}
\end{figure}

Both the current $I_z$ and flow rate $Q_z$ for $q>0$ clearly exhibit a nonlinear dependence on the electric field $E_z$.
Generally, the flow for $E_z<0$, the backward flow, is suppressed,
and completely blocked for $q=6q_0$.
For $E_z<0$, the colloidal particle is dragged close to the pore of $\mathcal{M}_1$ and serves as a valve that blocks the current and flow through the pore. A large colloidal particle charge enhances this effect.
To demonstrate the behavior of the colloidal particle, Fig.~\ref{fig:md_simulation} shows the probability distribution function for the colloidal particle along the $z$ axis for $q=6q_0$ and $E_z=\pm 0.7E_0$.
It is confirmed that the colloidal particle is mostly near membrane $\mathcal{M}_1$ under the backward electric field ($E_z=-0.7E_0$), whereas it is sufficiently far from $\mathcal{M}_1$ under the forward electric field ($E_z=0.7E_0$).

The MD results suggest that the nonlinear behavior in
the current and flow stems from the effective pore size being decreased by the colloidal particle.
We thus construct a theoretical model that captures the nonlinear responses, introducing an effective pore diameter that depends on the existence probability of the colloidal particle in the vicinity of the pore.
Using the effective pore diameter, we extend the theoretical model proposed by Sherwood et al.~\cite{MSG2014}, which gives the current and electro-osmotic flow induced in a cylindrical pore of finite length.
The model equations are given in the next section, followed by a comparison between the model and the results of the MD simulations in Sec.~\ref{sec:discussion}.
%

%
%
\section{\label{sec:model}Modeling of electro-osmotic diode }
In this section, we develop a mathematical model for the electro-osmotic diode described in the previous section to reproduce the MD results of Fig.~\ref{fig:result}.  
First, we outline the model equations for 
the current and electro-osmotic flow in a cylindrical pore of finite length
proposed by Sherwood et al.~\cite{MSG2014}, which we employ to express flows without colloidal particles.
We then develop a model equation for the effective pore diameter,
which depends on the existence probabilities of colloidal particles near the pore.
The Fokker--Planck equations governing the existence probabilities are summarized and the analytical solutions are derived.

%
%
%
\subsection{\label{subsec:theory} Electro-osmotic flow through nano-pore}
\begin{figure}[tb]
 \centering
 \includegraphics[width=0.95\hsize]{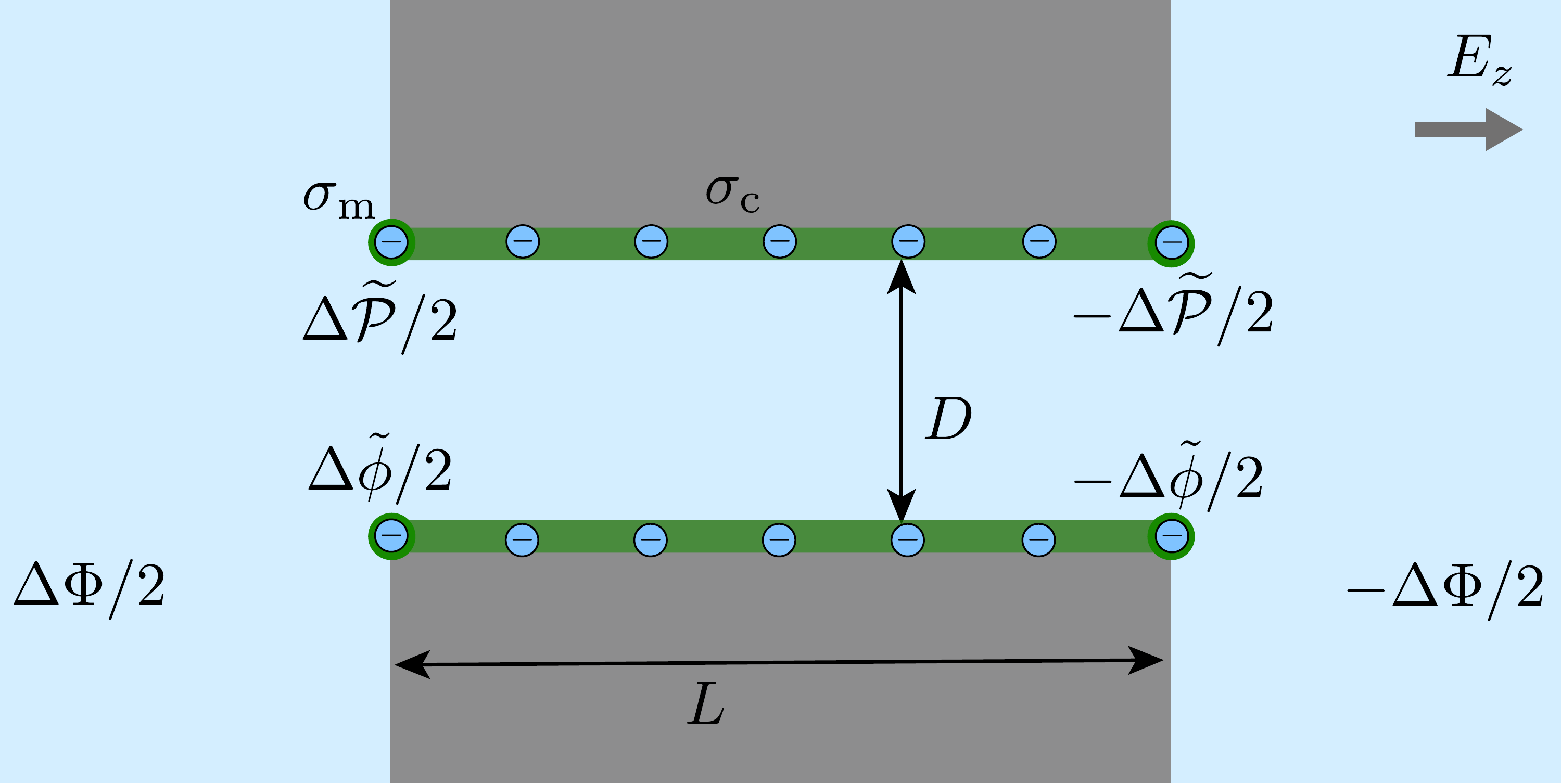}
 \caption{Schematic diagram of cylindrical pore of length $L$ and diameter $D$ considered in derivation of Eqs.~\eqref{eq:r_comp} and \eqref{eq:h_comp}.}
 \label{fig:eo-mem}
\end{figure}
We begin the model construction by considering 
the current and electro-osmotic flow through a cylindrical pore of finite length $L$ and diameter $D$ as shown in Fig.~\ref{fig:eo-mem}.
Here, we give a brief derivation of the model equations, the details of which are found in Refs.~\cite{SMG2014,MSG2014}.

In the linear response regime, the current $I_z$ and electro-osmotic flow rate $Q_z$ are written as follows:
\begin{align}
\label{eq:current}
I_z = - G \Delta\Phi,
\end{align}
\begin{align}
\label{eq:flow}
Q_z = - H \Delta\Phi,
\end{align}
where $G$ is the conductance, $H$ is the electro-osmotic flow coefficient,
and $\Delta \Phi$ is the electrical potential difference between the two sides of the membrane.
Before giving the explicit expressions for $G$ and $H$ for a finite-length cylindrical pore, we consider the extreme cases of a very thin membrane ($L\ll D$) and a thick membrane ($L\gg D$).

For $L\ll D$, the membrane is regarded as a zero thickness sheet with a pore, where the entrance effect is dominant.
The conductance is proportional to the diameter, namely $G_\mathrm{m} = \kappa D$, where $\kappa$ is the bulk electrical conductivity of the solution.
If we restrict ourselves to the small Debye length regime ($\lambda_\mathrm{d}\ll D$, as is true in the present MD simulations), the electro-osmotic flow coefficient is written as $H_\mathrm{m} \sim D\sigma_\mathrm{m} \lambda_\mathrm{d} /\mu$, where $\mu$ is the solution viscosity and $\sigma_\mathrm{m}$ is the charge density along the rim of the pore.
For a long cylindrical pore ($L\gg D$), the conductance is $G_\mathrm{c} = \pi D^2\kappa/(4L)$. For a small Debye length $\lambda_\mathrm{d}\ll D$, the electro-osmotic flow coefficient is written as $H_\mathrm{c} \sim \pi D^2 \sigma_\mathrm{c} \lambda_\mathrm{d}/(4\mu L)$, where $\sigma_\mathrm{c}$ is the surface charge density on the cylinder wall. 
Note that here and in what follows we use subscript $\mathrm{m}$ to represent quantities for the thin membrane with a pore ($L\ll D$), and subscript $\mathrm{c}$ for the long cylindrical pore ($L\gg D$).

We now give the expressions of $G$ and $H$ for a cylindrical pore of finite length $L$ using the coefficients for the limiting cases mentioned above.
The conductance $G$ is simply obtained by combining the effects of the entrance and the cylinder in series:
\begin{align}
\label{eq:r_comp}
G = \left(G_\mathrm{m}^{-1} + G_\mathrm{c}^{-1}\right)^{-1} =  \kappa \left(\dfrac{4L}{\pi D^2}+\dfrac{1}{D}\right)^{-1}.
\end{align}
In the expression of the electro-osmotic flow coefficient $H$, 
we need to take into account the effect of internal pressure difference $\Delta\widetilde{\mathcal{P}}$,
which is generated between the ends of the finite-length cylinder (see e.g. Ref.~\cite{YKW2014}).
For convenience in the following discussion,
we denote the electrical potential difference
between the two ends of the cylinder by $\Delta \tilde{\phi}$.
The internal values $\Delta\widetilde{\mathcal{P}}$ and $\Delta\tilde{\phi}$ are
obtained using the two continuity equations for the current and flow rate.
More specifically, the continuity of the current through the membrane is written as
\begin{equation}
    I_z=G_\mathrm{m}(\Delta\Phi-\Delta\tilde{\phi})=G_\mathrm{c}\Delta\tilde{\phi},
    \label{eq:continue1}
\end{equation}
where the left-hand side of the second equality is the current outside the membrane and the right-hand side is that inside the membrane.
The continuity of the flow rate $Q_z$ is written 
in terms of the flow rates generated by both the potential difference and the pressure difference~\cite{YKW2014}:
\begin{equation}
    Q_z=H_\mathrm{m}(\Delta\Phi-\Delta\tilde{\phi})-\mathcal{L}_m\Delta\widetilde{\mathcal{P}}
    =H_\mathrm{c}\Delta\phi+\mathcal{L}_\mathrm{c}\Delta\widetilde{\mathcal{P}},
    \label{eq:continue2}
\end{equation}
where $\mathcal{L}_m=-D^3/(24\mu)$ is Sampson's formula~\cite{Sampson1891} for the permeance of a pore on an infinitely thin sheet and $\mathcal{L}_c=-\pi D^4/(128L\mu)$ is Poiseuille's law
for the permeance of a cylindrical pipe.
The left-hand side of the second equality is the sum of the flow rates outside the membrane induced by the electrical potential and pressure differences,
and the right-hand side is that inside the cylinder.
Now the explicit expressions of $\Delta \tilde{\phi}$ and $\Delta \widetilde{\mathcal{P}}$ are given by
solving Eqs.~\eqref{eq:continue1} and \eqref{eq:continue2},
and the flow rate is then calculated using the first equality of Eq.~\eqref{eq:continue2} as follows:
\begin{align}
H&=
\dfrac{Q_z}{\Delta\Phi}
=\frac{\mathcal{L}_m H_\mathrm{c}  G_\mathrm{c}^{-1}+ \mathcal{L}_\mathrm{c}  H_\mathrm{m} G_\mathrm{m}^{-1}}{(\mathcal{L}_\mathrm{m} + \mathcal{L}_\mathrm{c})(G_\mathrm{m}^{-1} + G_\mathrm{c}^{-1})}
\\
&= \dfrac{
\left({\lambda_\mathrm{d}}/{\mu}\right)
\left(D\sigma_\mathrm{m} 
+ {16L}\sigma_\mathrm{c}/{3\pi}\right)
}{\left(1 + {4L}/{\pi D}\right)\left(1+{16L}/{3\pi D}\right)}.
\label{eq:h_comp}
\end{align}
%

%
\subsection{\label{subsec:model_eq}Effective pore diameter}
In our model of the electro-osmotic diode,
the change in the current and flow
is captured by introducing 
the effective pore diameter $D_\mathrm{eff}$,
which replaces $D$ in Eqs.~\eqref{eq:r_comp} and \eqref{eq:h_comp}
to capture the effect of the presence of a colloidal particle near the pore entrance.
We propose the following form for $D_\mathrm{eff}$:
\begin{align}
\label{eq:model_d}
D_\mathrm{eff} = D \exp \big(-\alpha \left(P_z(q,E_z) - P^0_{z}\right)
\left(P_{xy}(q) - P^0_{xy}\right) \big),
\end{align}
where $\alpha >0$ is a model parameter (constant), $P_z(q,E_z)$ is a quantity proportional to the probability that the colloidal particle
is close to membrane $\mathcal{M}_1$,
and $P_{xy}(q)$ is related to the probability that 
the colloidal particle is around the pore;
$P^0_z$ and $P^0_{xy}$ are the values at $q=0$.
Because $P_{xy}$ represents the colloidal particle motion in the $x$-$y$ plane, 
it is assumed to be independent of $E_z$.

In deriving this simple model expression,
we assume the small variation in $D_{\mathrm{eff}}$ is related to the probability $p^*$ (which represents the variables in the exponential in Eq.~\eqref{eq:model_d}) such that $\mathrm{d}p^*\sim -(D_{\mathrm{eff}}\mathrm{d}D_{\mathrm{eff}})/(\pi D_{\mathrm{eff}}^2)$,
i.e. $p^*$ is in proportion to the ratio of the variation in pore area.
Equation~\eqref{eq:model_d} is then obtained by integrating it under the requirement of $D_{\mathrm{eff}}$ at $p^*\to 0$.
Accordingly the essential feature of the diode is captured by this equation. 
If the colloidal particle is sufficiently far from $\mathcal{M}_1$
and thus $P_z \sim 0$, then the effective diameter is equal to the pore diameter ($D_\mathrm{eff} \sim D$).
In contrast, if the colloidal particle is near $\mathcal{M}_1$ ($P_z \gg 0$)  {\it and} around the pore ($P_{xy}\gg 0$), then the pore is completely clogged ($D_\mathrm{eff}\sim 0$).

In the following subsection, we discuss the analytical expressions for $P_z$ and $P_{xy}$ using the solutions of the Fokker--Planck equation. 

%
%
\subsection{\label{subsec:fp-eq}Fokker--Planck equation for colloidal particle motion}
Here, we discuss the motion of the colloidal particle employing the Fokker--Planck equations to obtain the expressions for the parameters $P_{z}$ and $P_{xy}$, which are included in the effective diameter.
The colloidal particle in the solution under an electrical field is subjected to the electro-phoretic force and random forces from the solvent particles,
which leads to advection and diffusion. 
In the following, the Fokker--Planck equations governing the probability density under this situation
are separately formulated for motion along the $z$ axis and in the $x$-$y$ plane.

The Fokker--Planck equation for the probability density function $p_z$ is
\begin{align}
\label{eq:fp}
\frac{\partial p_z}{\partial t} =\frac{\partial}{\partial z}\left[(Ap_z) + \frac{\partial}{\partial z}(\mathcal{D}p_z) \right],
\end{align}
where the coefficient $A(q,E_z)$ corresponds to advection and 
$\mathcal{D}$ is the diffusion coefficient for the particle, which we assume to be independent of $q$ and $E_z$.
In the steady-state, the analytical solution is written as 
\begin{align}
& p_z(z) = p_0 \exp\left(-\dfrac{Az}{\mathcal{D}}\right),
\label{eq:ana_sol}\\
& p_0 = {A}\mathcal{D}^{-1}\left[\exp\left(-\frac{A}{\mathcal{D}}z_1\right) - \exp\left(-\frac{A}{\mathcal{D}}z_2\right)\right]^{-1},
\label{eq:ana_sol_p0}
\end{align}
where $z_1$ and $z_2$ are the positions of boundaries at the surface of $\mathcal{M}_1$ and $\mathcal{M}_2$, respectively (see dashed lines in Fig.~\ref{fig:md_simulation}).

Next, we focus on the motion of the colloidal particle in the $x$-$y$ direction. We consider the existence probability distribution $p_r$ in the cylindrical coordinate system.
The Fokker-Planck equation is written as follows:
\begin{align}
\label{eq:fp_polar}
\frac{\partial p_r}{\partial t} = \frac{1}{r}\dfrac{\partial}{\partial r}\left(A_r rp_r \right) 
+ \mathcal{D}\left[\frac{1}{r}\dfrac{\partial}{\partial r}\left(r\dfrac{\partial}{\partial r}\right)\right]p_r.
\end{align}
Here, we assume that coefficient $A_r$ is inversely proportional to $r^2$, i.e., $A_r(q,r)={C_A(q)}/{r^2}$, to take into account radial advection due to electrical interaction between the charged particle and the pore. Then, the following analytical solution is obtained:
\begin{align}
&p_r(r) = p_{r0} \exp\left(\frac{C_A}{\mathcal{D}r}
\right),\label{eq:pr_ana}
\\
&p_{r0}=\left[
\int_0^{r_\mathrm{max}} r\exp\left(\frac{C_A}{\mathcal{D}r}\right)\mathrm{d}r
\right]^{-1}.\label{eq:pr0_ana}
\end{align}

The parameters in the effective diameter
in Eq.~\eqref{eq:model_d} are defined as
$P_z=p_z(z_1)$ and $P_{xy}=\int_0^{\delta r}r' p_r(r')\mathrm{d}r'$
where $\delta r$ is the radius of the vicinity region around the pore.
We discuss the actual values of these parameters
corresponding to the MD setup in Sec.~\ref{sec:problem} with the aid of independent MD simulations conducted to estimate parameters.

%
\section{\label{sec:discussion}Comparison of model with MD simulations}

%
\subsection{\label{subsec:param_id} Model parameters}
Here, we determine the actual values of the model parameters for the system used for our MD simulations.
We first estimate the diffusion coefficient $\mathcal{D}$ for the colloidal particle in the solution. To this end, we perform independent MD simulations for Brownian motion of the colloidal particle in the bulk electrolyte solution (see Supplemental Information S1 for details).
From the mean square displacement of the particle, 
the diffusion coefficient is obtained as $\mathcal{D}=1.7\times 10^{-2} \sigma_0^2/\tau_0$.
Because the colloidal particle charge has little effect on diffusive motion, we use this value throughout the following discussion.

We next consider the effect of advection in the $z$ direction, which is incorporated in the model via the parameter $A(q,E_z)$ in Eqs.~\eqref{eq:ana_sol} and \eqref{eq:ana_sol_p0}.
The motion of the colloidal particle with various values of charge $q$ is tracked using an MD simulation setup where the colloidal particle is placed between two virtual membranes (same as $\mathcal{M}_2$ used in Sec.~\ref{sec:problem}) with an electric field $E_z$ applied. (see Supplemental Information S2 for details). 
We obtain the probability density function for the colloidal particle along the $z$-axis, and determine the values of the coefficient $A$ at each $(q,\,E_z)$ by fitting the analytical solution given by Eqs.~\eqref{eq:ana_sol} and \eqref{eq:ana_sol_p0} using the value of $\mathcal{D}$ obtained above.
We found the form $A = \beta qE_z$ with a constant $\beta =-1.3\times 10^{-2}\sigma_0^2/\epsilon_0\tau_0$.

Finally, we examine the effect of advection in the $x$-$y$ direction
to determine the value of $C_A$ in Eqs.~\eqref{eq:pr_ana} and \eqref{eq:pr0_ana}.
Here, we consider the situation where the motion of the colloidal particle is 
constrained near membrane $\mathcal{M}_1$ (see Supplemental Information S3 for details).
This constraint is realized by placing the virtual membrane (same as $\mathcal{M}_2$)
at a distance of one diameter of the colloidal particle away from $\mathcal{M}_1$.
To focus on the motion in the $x$-$y$ plane, the pore charges are distributed as in Sec.~\ref{sec:problem}, but the physical pore is not taken into account.
Using the same value for $\mathcal{D}$, we determine the value of $C_A$ by fitting the analytical solution given by Eqs.~\eqref{eq:pr_ana} and \eqref{eq:pr0_ana} to the existence probability density functions obtained with these constrained MD simulations.
Accordingly, the value of $C_A$ was obtained in the form $C_A=\gamma q$, with $\gamma=6.2\times 10^{-4}\sigma_0^3/\tau_0q_0$.

\begin{figure}[t]
 \centering
 \includegraphics[width=0.95\hsize]{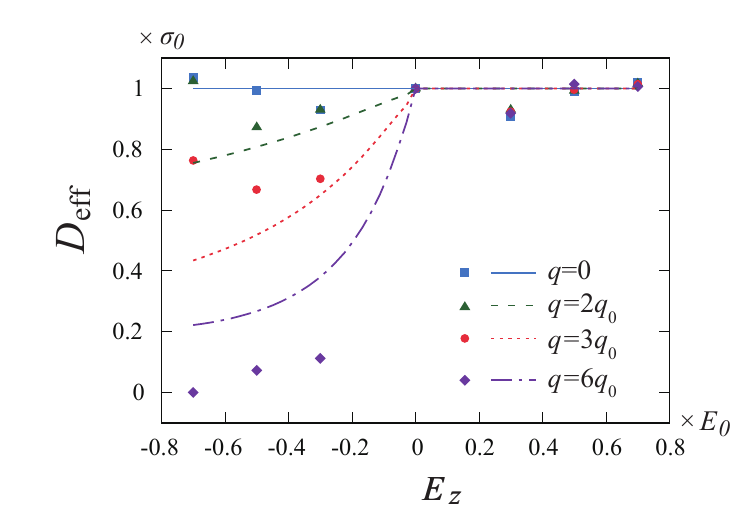}
 \caption{Effective diameter of nano-pore $D_\mathrm{eff}$ as a function of electric field $E_z$ for various values of colloidal particle charge $q$. The symbols indicate the values estimated from the full MD simulation and the lines indicate the model given in Eq.~\eqref{eq:model_d}.}
 \label{fig:d_eff}
\end{figure}
%
%
\subsection{\label{subsec:comparison}Comparison with full MD simulation}

With the obtained values of the model parameters $\mathcal{D}$, $A$, and $C_A$
from independent MD simulations, we are able to estimate the effective diameter of the pore in the presence of the colloidal particle.
The current and electro-osmotic flow estimated by the model given in Eqs.~\eqref{eq:current}, \eqref{eq:flow}, \eqref{eq:r_comp}, and \eqref{eq:h_comp} using the effective pore diameter $D_\mathrm{eff}$ defined in Eq.~\eqref{eq:model_d} are shown in Fig.~\ref{fig:result}.
Here, the free parameter $\alpha$ is set to $\alpha=6.7$.
Both the current and electro-osmotic flow rate are in good agreement with the full MD simulation results, showing correct nonlinear behavior of an electro-osmotic diode. 
The results indicate that with appropriate values of model parameters, the performance of the electro-osmotic diode can be approximately estimated from the model equation as a function of the charge of the colloidal particle $q$ and electric field $E_z$.

A direct comparison of the value of the effective diameter $D_\mathrm{eff}$ is shown in Fig.~\ref{fig:d_eff}. Here, the values of $D_\mathrm{eff}$ for the full MD simulations are inversely computed from the MD results in Fig.~\ref{fig:result} using Eqs.~\eqref{eq:current}, \eqref{eq:flow}, \eqref{eq:r_comp}, and \eqref{eq:h_comp} with the diameter $D$ as a unknown parameter.
Good agreement between the MD simulations and the model equation is obtained.

The other physical parameters in the model are set as follows. Parameters such as the surface charge densities $\sigma_\mathrm{c}$ and $\sigma_\mathrm{m}$ were determined from the geometrical setup.
The electrical conductivity $\kappa$ and viscosity $\mu$, which are related to the properties of the electrolyte solution, could also be measured independently. 
In the comparison above, however, to bypass the independent measurements of these parameter values, we obtained them from a linear approximation of the current and flow rate in the region $E_z\geq 0$, where the pure current and pure electro-osmotic flow are approximately obtained because the colloidal particle is absent near the pore.
%
%

\section{\label{sec_conclusion}Concluding remarks}
In this study, we proposed an electro-osmotic diode that consists of colloidal particles between two membranes.
The difference in pore density between two membranes makes the colloidal particles serve as {\it nano-valves}. The rectification of the current and electro-osmotic flow is observed in MD simulations of the entire system. We established an analytical model that captures the key feature of the electro-osmotic diode, i.e., the colloidal nano-valve, where a colloid particle clogs the pore to prevent the backward current and flow.

The behavior of the colloidal nano-valve is modeled employing the introduced effective pore diameter, which depends on the existence probability density functions around the pore, as defined in Eq.~\eqref{eq:model_d}.
With the estimated model parameters, we obtained the forms of the probability density functions using the analytical expressions for the Fokker--Planck equations for the advection and diffusion of a particle. 
In our comparison with the full MD simulations, we estimated the parameter values using several independent MD simulations of colloidal motion, namely a bulk simulation to determine the diffusion coefficient, and two simulations for a colloidal particle in closed domains to determine the advection parameters in the $z$ and $x$-$y$ directions. 
Using the obtained parameter values, the full MD simulation results are successfully reproduced by the model equation, as shown in Figs.~\ref{fig:result} and \ref{fig:d_eff}.
This confirms that the proposed model reproduces the essential mechanism of the colloidal nano-valve. The quantification of the free parameter $\alpha$ is obtained via a top-down estimation in the comparison. 
Though one-time calibration of $\alpha$ is required for direct comparison,  investigation of the diode performance for various situations becomes possible once the value of $\alpha$ is identified. Further investigation into the microscopic colloidal particle motion near the pore would allow us to determine the value of $\alpha$ using a bottom-up approach. This topic will be considered in future work. 
We also note here that this idea of effective pore diameter has other potential applications, such as molecular sieving, nano-filtration, and transports in porous media as in liquid electrolyte secondary batteries.

Throughout the paper, the membrane with a high number density of pores (membrane $\mathcal{M}_2$) is regarded as a complete semi-permeable membrane,
in which only the colloidal particle is affected by the potential barrier. 
This assumption was made to focus on the geometrical constraint of the colloidal particle. 
In a real membrane, however, the electro-osmotic flow is also driven in the pores in $\mathcal{M}_2$. This effect might not be negligible, and it would greatly enhance the magnitude of the current and the electro-osmotic flow while maintaining the basic mechanism of the nano-valves.
Studies with the explicit configuration of $\mathcal{M}_2$ would thus reveal this flow enhancement.
Furthermore, the assumption of abstract materials for the colloidal particles, membranes, and electrolyte solution, can be replaced by the adoption of practical materials. Taking into account real materials, along with the design of appropriate experimental setups, is also a future research topic.

The proposed electro-osmotic diode is constructed with combining different physics, namely the electro-osmotic flow occurring in nano/micro pores, electrophoretic motion of colloidal particles, and the mechanical valve effect. We hope this crosscutting idea, including the mathematical modelling, will promote interactions among scientists in different fields. On the practical side, the present diode has potential application in rectifying flows under an AC electric field if the period is sufficiently longer than the characteristic relaxation time of the rectification. The {\it nano-valves} allow asymmetric flow and block backward flow almost completely if the physical parameters are appropriately chosen. 
Therefore, more efficient rectification compared to that achieved by existing AC electro-osmotic systems is expected. 
We hope that the present simple configuration of the diode system will accelerate the development of practical electro-osmotic pumps in micro- and nano-fluidic devices.

%
\section*{Acknowledgments}
This research was partially supported by Intelligent Mobility Society Design, Social Cooperation Program (Next Generation Artificial Intelligence Research Center, The University of Tokyo, and Toyota Central R\&D Labs., Inc.)
This project was granted access to HPC facilities financed by Initiative on Promotion of Supercomputing for Young or Women Researchers, Information Technology Center, The University of Tokyo.
\section{appendix}
\appendix
\section{\label{sm:bulk} Diffusive motion of a colloidal particle}
\begin{figure}[t]
 \centering
 \includegraphics[width=0.7\hsize]{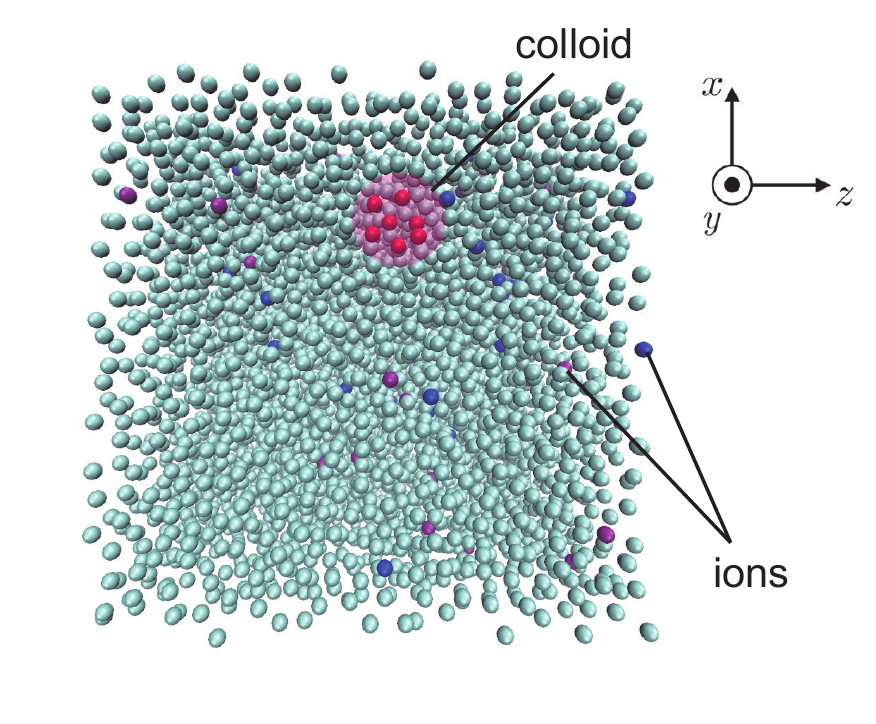}
 \caption{The system of the bulk MD simulation. The system consists of a colloidal particle and solution. The periodic boundary conditions are assumed in the all directions.}
 \label{fig:md_bulk}
\end{figure}

The bulk MD simulations are performed to investigate the diffusive motion of a colloidal particle. 
The colloidal particle is put in a cube simulation box as shown in Fig.~\ref{fig:md_bulk}. 
The system contains $6290$ solution particles including $50-q$ monovalent cations and $50$ monovalent anions. The initial configuration is equilibrated such that the pressures for all directions are at $1 \pm 0.1 \epsilon_0/\sigma_0^3$. The colloidal particle with charge $q$ has the same structure as that of Fig.~1 in the main text.
No electric field is applied. The colloidal particle charge is set at $q\,(=6,\,3,\,2,\,0\,q_0)$, which are used in the full MD simulations in Fig.~2.

\begin{figure}[tb]
 \centering
 \includegraphics[width=0.75\hsize]{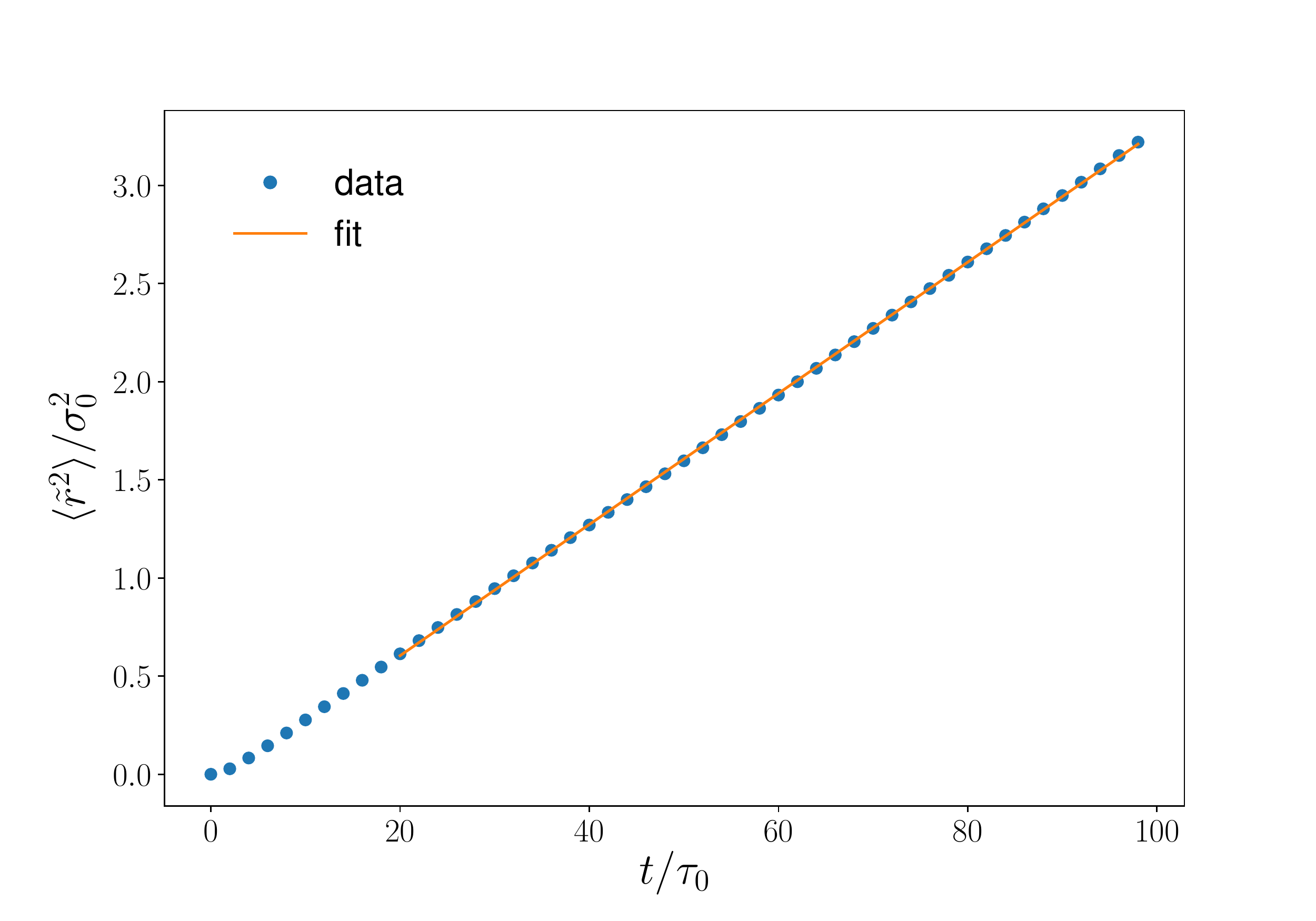}
 \caption{The mean square displacement (MSD) versus time obtained from the bulk simulations, for the case of $q=6\,q_0$.
 The diffusion coefficient is calculated from slope between $t = 20 \tau_0$ and $100 \tau_0$.}
 \label{fig:msd}
\end{figure}

After long time has passed, the mean square displacement (MSD) $\langle \tilde{r}^2 \rangle=({1}/{3})\langle \bm{x}^2 \rangle$ becomes proportional to time $t$, which is a typical feature of diffusive motion. The MSD is then written in the form $\langle \tilde{r}^2 \rangle=2\mathcal{D}t$ with $\mathcal{D}$ being the diffusion coefficient.
The plot of the MSD for the cases of $q=6\,q_0$ is shown in Fig.~\ref{fig:msd}. The diffusion coefficient $\mathcal{D}$ is obtained from linear fitting in the long-time regime. 
Practically, we used the data for $t \gg 20\tau_0$, which is sufficiently larger than the mean free time.
The colloid charge had little impact, and the mean value of the diffusion coefficient was obtained as $\mathcal{D}= 1.72\times 10^{-2}\pm 0.09\, \sigma_0^2/\tau_0$. The simulations were performed for $10^7$ steps with the time step $dt=0.002\tau_0$. Three runs with different initial configurations were used for each $q$. In obtaining the MSD for each run, samples of for the time window of $100\tau_0$ ($5 \times 10^4$ steps) were averaged with shifting the window by $1000$ steps.
%
\section{\label{sm:nw} Advection of a colloidal particle along the electric field}
\begin{figure}[tb]
 \centering
 \includegraphics[width=0.95\hsize]{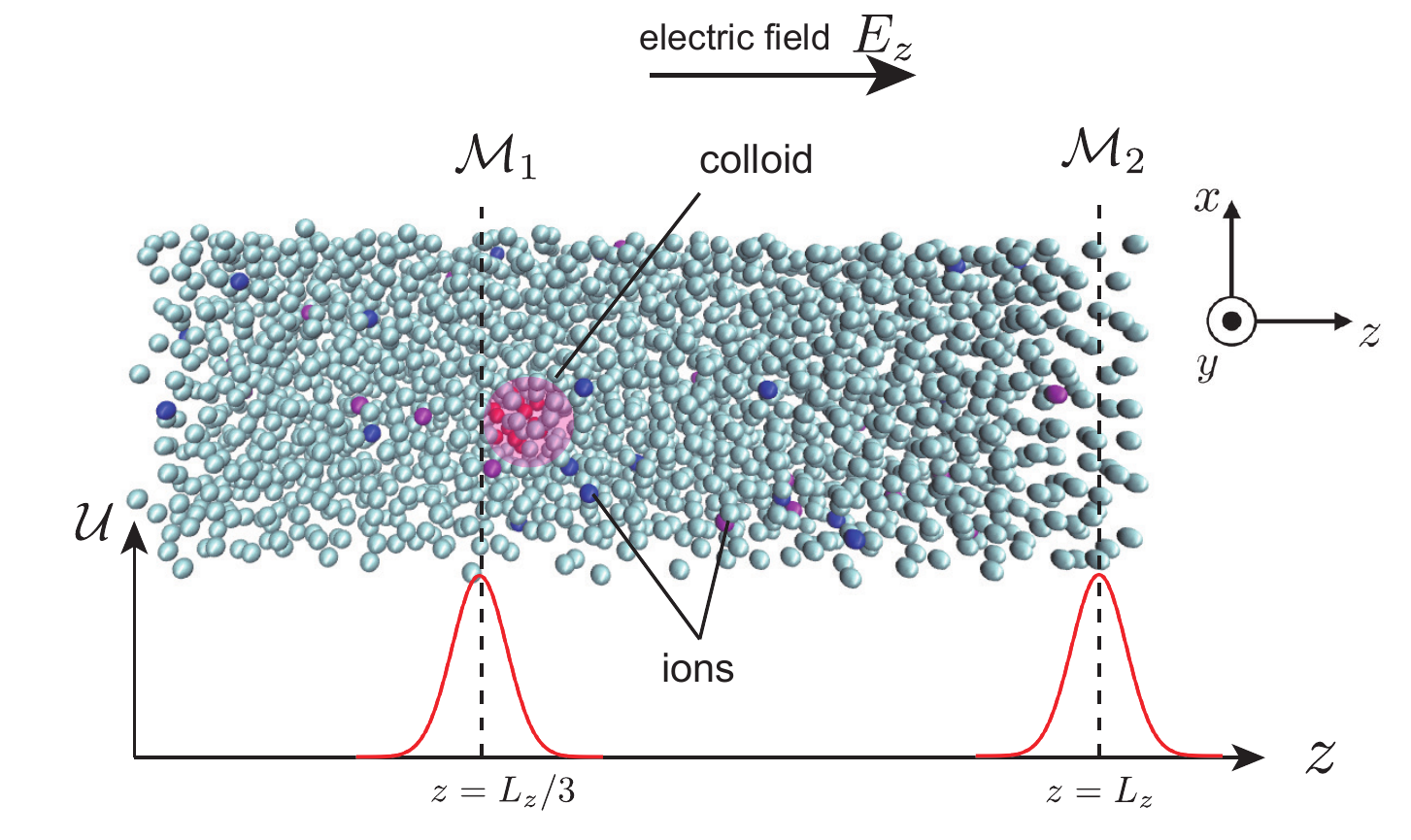}
 \caption{The system of MD simulations. A colloidal particle is put between two potential barriers $\mathcal{M}_1$ and $\mathcal{M}_2$.}
 \label{fig:md_nw}
\end{figure}

The coefficient for the advection in the Fokker--Planck equation ($A$ in the main text) is evaluated using the system in which both membranes $\mathcal{M}_1$ and $\mathcal{M}_2$ are the virtual membrane of potential barriers as shown in Fig.~\ref{fig:md_nw}.
The colloidal particle with the charge $q$ has the same structure as in Fig.~1.
The system contains $2410$ solution particles including $20-q$ monovalent cations and $20$ monovalent anions, and the system size is $L_x=L_y=10$\,$\sigma_0$, and $L_z\sim 30\,\sigma_0$. Here, as in the full MD simulation (Fig.~1), the system is equilibrated such that the pressure in the $z$ direction $\mathcal{P}_z$ is $1.0 \pm 0.1 \epsilon_0/\sigma_0^3$. 
The periodic boundary condition is assumed in all directions.
The simulation system is illustrated in Fig.~\ref{fig:md_nw}.
Three production runs with different initial configurations are performed for $q=6,\,3,\,2\,q_0$.
At each production run, $10^7$ time-step simulation is carried out with the time step $dt=0.002\tau_0$. 

From the trajectories of the MD simulations, the existence probability $p_z(z)$ of the colloidal particles are computed for the steady state.
The applied electric field is varied in $-0.7\leq E_z \leq0.7\,E_0$ as in the main text. 
The distribution functions $p_z(z)$ are normalized such that $\int_{z_1}^{z_2}p_z(z')dz'=1$, where $z_1$ and $z_2$ are the effective membrane surface,
such that the positions of peaks of $p_z(z)$ coincide with the effective membrane surfaces.
\begin{figure}[tb]
    \centering
    \includegraphics[clip, width=0.7\columnwidth]{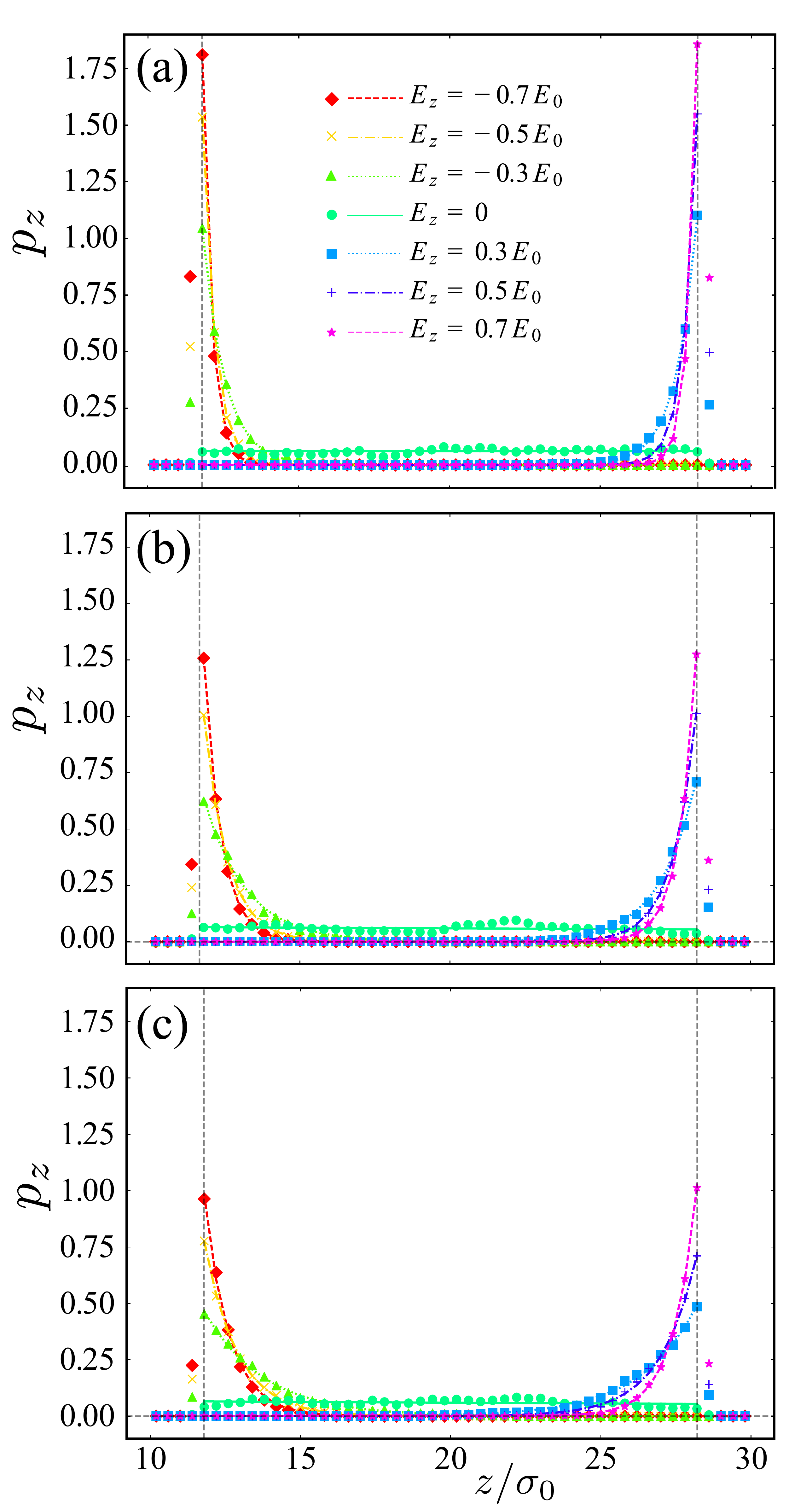}
    \caption{The existence probability density function of the colloidal particle along the $z$ axis for various values of the colloidal particle charge (a) $q=6\,q_0$,\, (b) $q=3\,q_0$ and (c) $q=2\,q_0$. The symbols and lines represent MD simulation data and fitted curve, respectively. 
    The electric field is varied from $-0.7$\,$E_0$ to $0.7$\,$E_0$.}
    \label{fig:pz_fit}
\end{figure}

\begin{figure}[t]
    \centering
    \includegraphics[width=1\hsize]{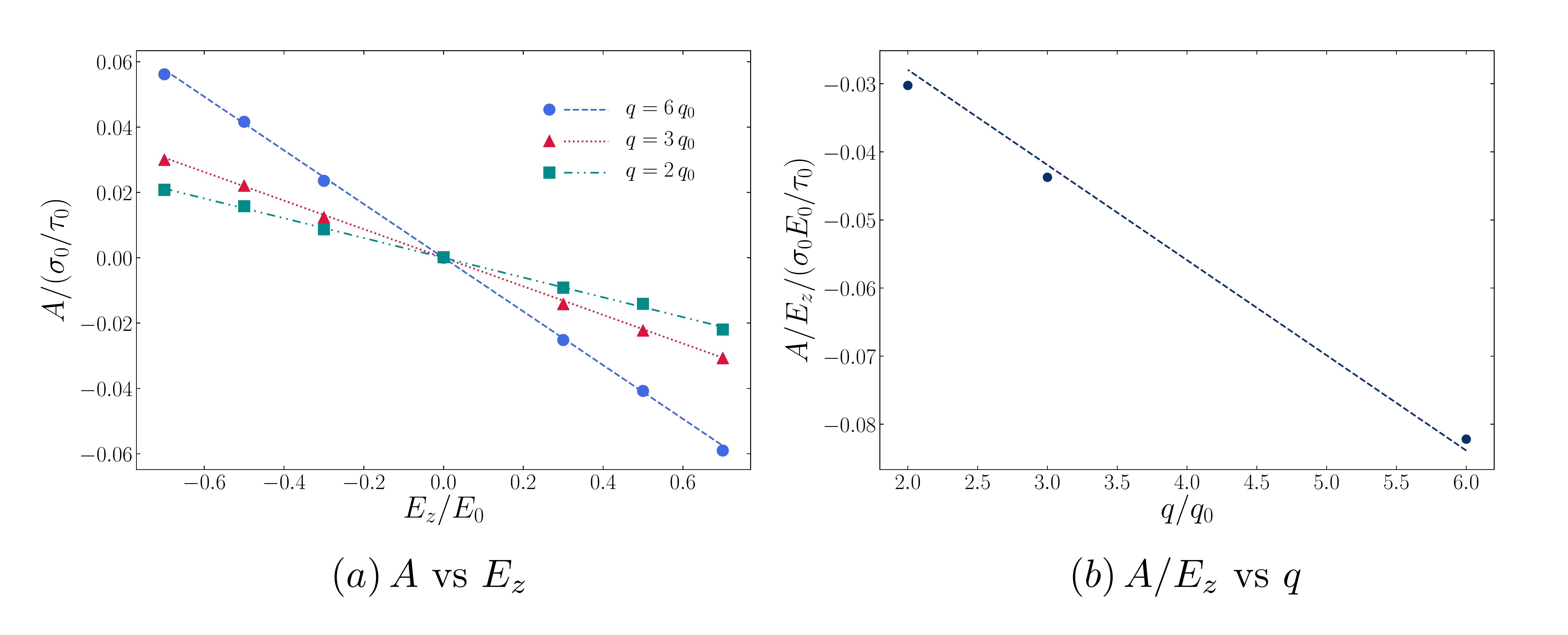}
    \caption{Advection coefficient obtained from the MD simulation (symbols) with linear fit (lines). (a) $A$ vs $E_z$ and (b) $A/E_z$ vs $q$.}
    \label{fig:advection}
\end{figure}

\begin{figure}[tb]
 \centering
 \includegraphics[width=0.75\hsize]{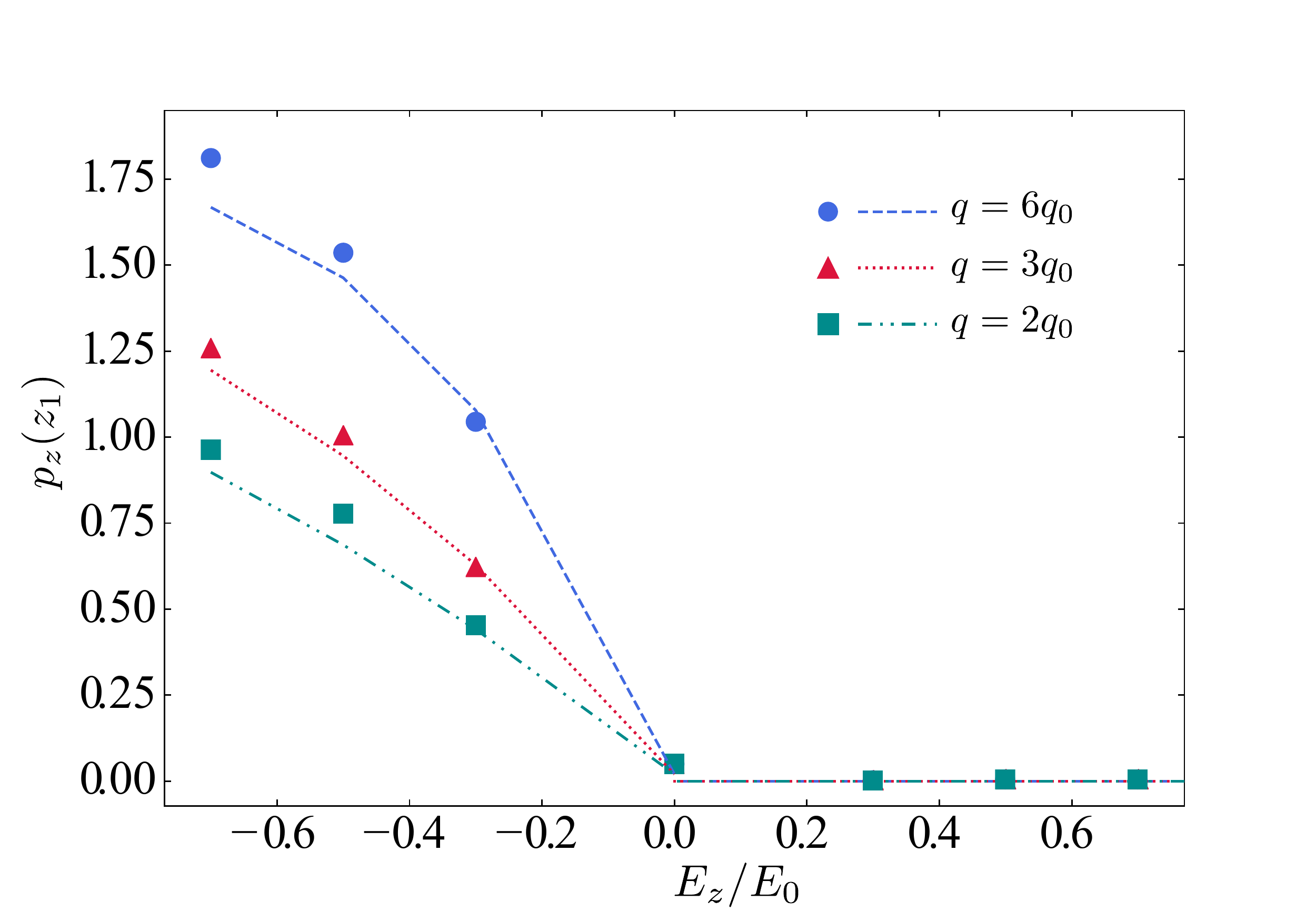}
 \caption{The comparison between simulation data (symbols) and analytic values (lines). The values of $p_z(z_1)$ are plotted for three values of colloid charge $q$.}
 \label{fig:p_max}
\end{figure}

The advection coefficient $A$ is obtained using the numerically measured $p_z(z)$, by fitting the analytical solution of the Fokker--Planck equations Eqs.~(11) and (12) with 
 the least-squares method. The results are shown in Fig.~\ref{fig:pz_fit}.
 Here, the diffusion coefficient $\mathcal{D}$ is fixed at $1.72\times 10^{-2}\,\sigma_0^2/\tau_0$, which was obtained in the Sec.~\ref{sm:bulk}.
The obtained values of the advection coefficient $A$ are plotted as a function of the electric field $E_z$ in Fig.~\ref{fig:advection}(a).
As shown by the fitted linear lines, the dependence on $E_z$ is linear at each value of $q$.
Next, the slopes of the fitted linear lines are plotted as a function of $q$
in Fig.~\ref{fig:advection}(b), which is also linear.
Altogether, we infer the form of function $A(q,E_z) = \beta qE_z$, 
The coefficient is evaluated as $\beta=-1.3\times10^{-2}\sigma_0^2/\epsilon_0\tau_0$ from the slope of (b).
Using this analytical form of $A(q,E_z)$ together with the value of $\mathcal{D}$, we can  now obtain the analytical value of the existence probability at the surface of $\mathcal{M}_1$, $p_z(z_1)$, which is used in the definition of the effective pore diameter.
In Fig.~\ref{fig:p_max}, we compare the analytically obtained $p_z(z_1)$ and the simulation results (symbols), showing a good agreement.
%
\section{\label{sm:polar} Motion of a colloidal particle in parallel to the membrane}

\begin{figure}[tb]
 \centering
 \includegraphics[width=0.9\hsize]{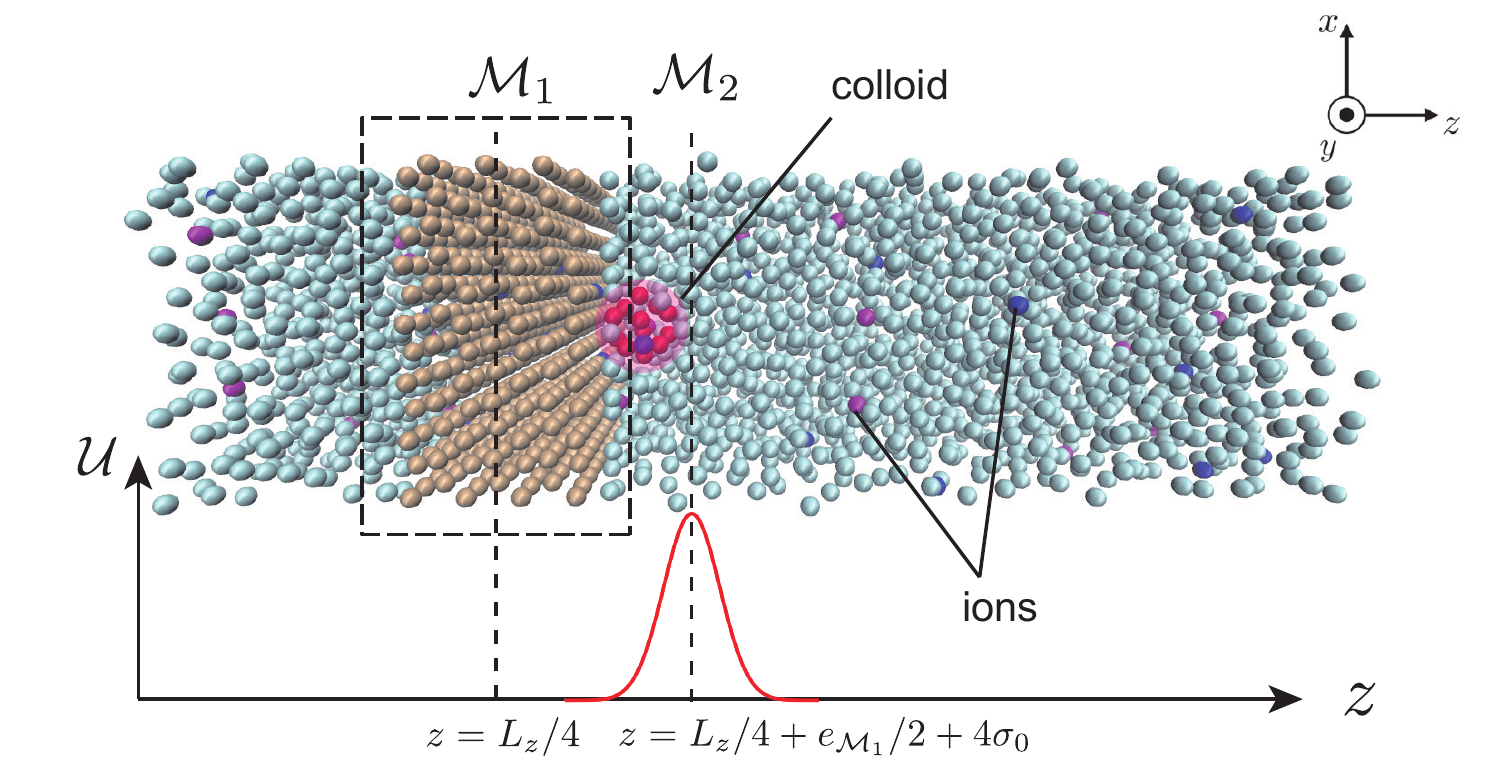}
 \caption{The system of MD simulations dedicated to evaluating parameters related to the colloidal motion in $x$-$y$ plane. System consists of a colloidal particle between physical membrane $\mathcal{M}_1$ and potential barrier $\mathcal{M}_2$ and solution. The interval between $\mathcal{M}_1$ and $\mathcal{M}_2$ is set to the extent of the colloidal particle diameter, such that particle motion is restrained in $x-y$ plane.}
 \label{fig:md_polar}
\end{figure}

Here, we discuss the existence probability of a colloidal particle close to the membrane.
To this end, we consider a system with plane membrane as $\mathcal{M}_1$, and a virtual energy barrier $\mathcal{M}_2$ that is placed near $\mathcal{M}_1$ to restrict the motion of colloidal particles to the region close to the membrane.
Specificaly, the position of $\mathcal{M}_2$ is at $z=L_z/4+e_{\mathcal{M}_1}/2 + 4\sigma_0$, where $e_{\mathcal{M}_1}$ is the width of $\mathcal{M}_1$.
In order to prevent the colloidal particle from being trapped in the pore, $\mathcal{M}_1$ has no pore, while the pore charges are distributed in the same manner as in the main text.
The numbers of solvent and ion particles are the same as in the simulation of Fig.~1. The system used here is shown in Fig.~\ref{fig:md_polar}.

\begin{figure}[b]
 \centering
 \includegraphics[width=0.9\hsize]{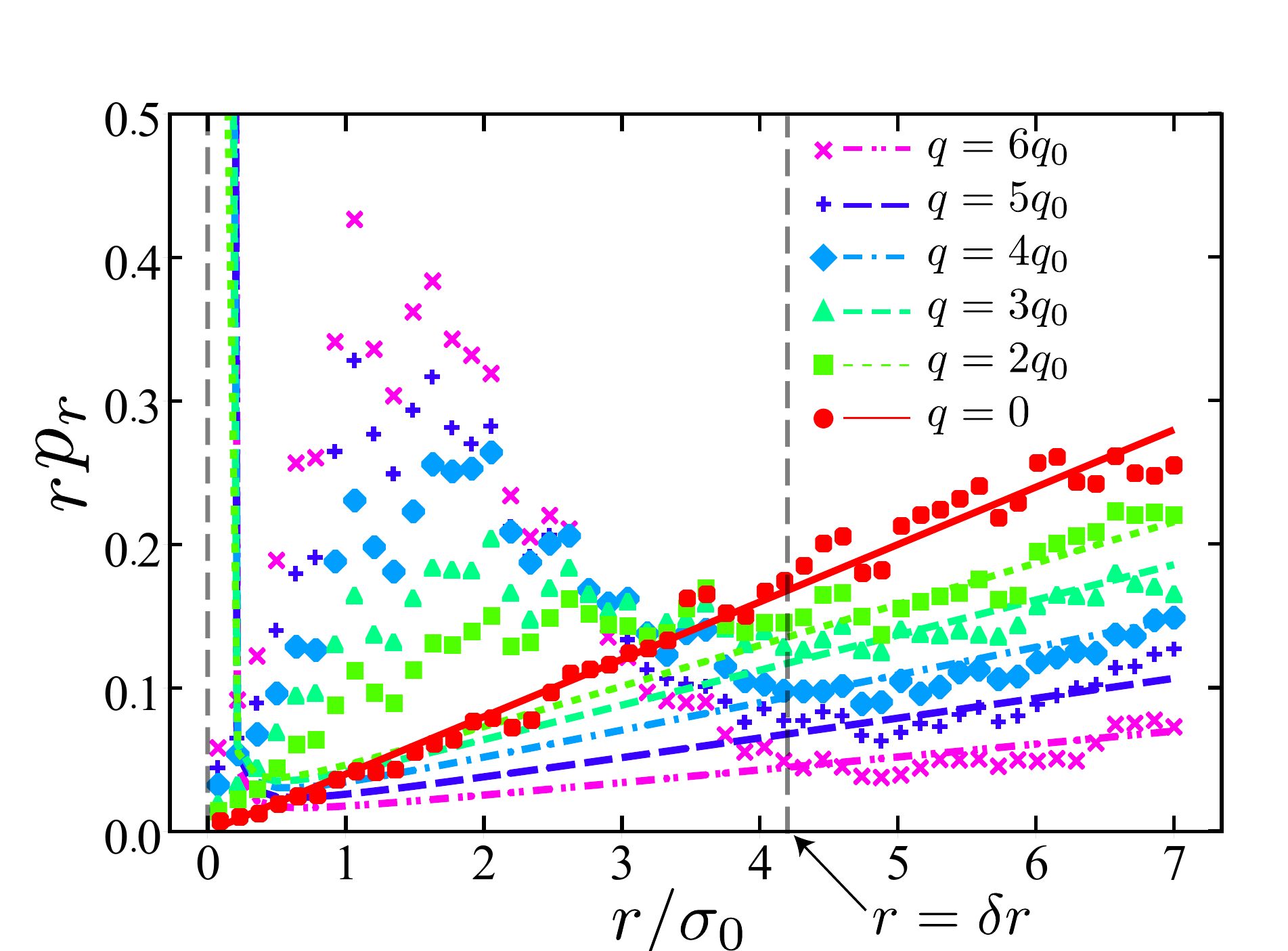}
 \caption{The comparison between simulation data (symbols) and analytic values (lines) for $p_r$. Analytic values are obtained using optimization with Eq.~\eqref{smeq:optimize}.}
\label{fig:pr_fit}
\end{figure}

From the MD simulations we compute the distribution of colloidal particles in the $r$ direction $p_r(r)$, where $r$ is the radial coordinate in the cylindrical coordinate system in the plane parallel to the membrane. 
The tail of the probability distribution is appropriately corrected using the nature of the periodic boundary conditions.
We plot the measured $p_r(r)$ in Fig.~\ref{fig:pr_fit}. 
As in the previous section, three production runs with different initial configurations are performed for $q=6,\,3,\,2\,q_0$.
At each production run, $10^7$ time-step simulation is carried out with the time step $dt=0.002\tau_0$. 

We compare the results of MD simulation data with the analytical solution in Fig.~\ref{fig:pr_fit}. 
Note that the analytical solution is based on the negative point charge at the center of the membrane, while in the MD simulation the negative charges on the membrane are distributed in a finite region.
Therefore, to pot weight on the tail (large $r$) in fitting the analytical functions, we consider the following objective function $f$:
\begin{align}
\label{smeq:optimize}
f(r) = \left|\int_0^{\delta r} r p_r dr - \int_0^{\delta r} r h(r)dr\right|\\ \nonumber
+ w \sum_{r_i>\delta r}^{r_\mathrm{max}} | r_ip_r(r_i)-r_ih(r_i)|^2,
\end{align}
where $h(r)$ is the values of the histogram obtained from the simulations, while $p_r$ is the analytical solution; $\delta r \sim 4 \sigma_0$ is a threshold, which is about the radius of the region where the direct effects of actual negative charges are significant in the simulation.
The first term on the right-hand side is the difference in integrals of the distributions for $r<\delta r$, and the second term is the difference at each point of $r\geq \delta r$.
Putting emphasis on the tail ($r\geq \delta r$) of the distribution, we set a weight of $w=10$. We used the downhill simplex method for the optimization.
From the result of fitting, we found the dependence of $C_A$ on $q$ to be linear, i.e., $C_A = \gamma q$.  The value of the parameter $\gamma$ was found to be $\gamma=6.2\times 10^{-4}\,\sigma_0^3/\tau_0 q_0$.

%
\end{document}